\documentstyle[11pt,epsf]{l-aa}

\def\etal{{\it et al.~}}
\def\eg{{\it e.g.,~}}
\def\ie{{\it i.e.,~}}

\begin{document}

\thesaurus{ }

\title{Cosmic Magnetic Fields in Large Scale Filaments and Sheets}

\author{Dongsu Ryu$^{1,2,5}$, Hyesung Kang$^{1,3,6}$ and Peter L.
Biermann$^{4,7}$}

\institute{
$^1$Department of Astronomy, University of Washington,
    Box 351580, Seattle, WA 98195-1580, USA\\
$^2$Department of Astronomy \& Space Science, Chungnam National
    University, Daejeon 305-764, Korea (present address)\\
$^3$Department of Astronomy, University of Minnesota, Minneapolis,
    MN 55455, USA (present address)\\
$^4$Max-Planck Institute for Radioastronomy, Bonn, Germany\\
$^5$e-mail: ryu@canopus.chungnam.ac.kr\\
$^6$e-mail: kang@canopus.chungnam.ac.kr\\
$^7$e-mail: plbiermann@mpifr-bonn.mpg.de}
\offprints{Dongsu Ryu:\\ ryu@canopus.chungnam.ac.kr}

\date{received date; accepted date}
\maketitle

\markboth{Cosmic Magnetic Fields}{c}

\begin{abstract}

We consider the possibility that cosmic magnetic field, instead of being
uniformly distributed, is strongly correlated with the large scale structure
of the universe.
Then, the observed rotational measure of extra-galactic radio sources
would be caused mostly by the clumpy magnetic field in cosmological
filaments/sheets rather than by a uniform magnetic field, which was often
assumed in previous studies.
As a model for the inhomogeneity of the cosmological magnetic field, we adopt
a cosmological hydrodynamic simulation, where the field is passively included,
and can approximately represent the real field distribution with an arbitrary
normalization for the field strength. Then, we derive an upper limit of the
magnetic field strength by comparing the observed limit of rotational measure
with the rotational measure expected from the magnetic field geometry in the
simulated model universe. The resulting upper limit to the magnetic field in
filaments and sheets is ${\bar B}_{fs} \la 1 \mu{\rm G}$ which is $\sim10^3$
times higher than the previously quoted values.
This value is close to, but larger than, the equipartition magnetic field
strength in filaments and sheets.
The amplification mechanism of the magnetic field to the above strength
is uncertain.
The implications of such a strength of the cosmic magnetic field are
discussed.


\end{abstract}

\section{Introduction}

The strength and morphology of the intergalactic magnetic fields
remain largely unknown, because it is intrinsically difficult to observe
them (for recent reviews, see Kronberg 1994; Beck et al. 1996; Zweibel \&
Heiles 1997). While the magnetic fields inside typical galaxies are observed to
be of order of $3-10\mu{\rm G}$, recent observations indicate that
the magnetic fields of the similar strength are also common in core regions
of rich clusters (for a recent discussion see En{\ss}lin \etal 1997).
On the other hand, the upper limit for a large-scale field strength placed
by the observed rotation measure (RM) of quasars is about
$B_{IGM} \la 10^{-9} L_{\rm rev,Mpc}^{-1/2}{\rm G}$ (Kronberg 1994).
Here, $L_{\rm rev,Mpc}$ is the reversal scale of the magnetic fields in
units of Mpc in comoving coordinates.
It was assumed the fields are uniform in direction within the reversal
scale and varied as a comoving passive field in the expansion of the
universe. The above value is close to the limit placed by the observed 
anisotropy in the cosmic microwave background radiation is 
$B_{IGM} < 6.8 \times 10^{-9} (\Omega_o h^2)^{1/2} {\rm G}$
(Barrow, Ferreira \& Silk 1997).
Here, a uniform magnetic field was assumed.
$h$ is $H_o$ in unit of $100~{\rm km/s/Mpc}$.

According to a recent popular view based on both observational and
theoretical cosmology, the dominant nonlinear structure
in the universe is a web-like network of filaments
(\eg Bond, Kofman \& Pogosyan 1996; Shectman \etal 1996).
At a lower density contrast, however, bubbly walls whose
intersections are in fact filaments are the dominant structures.
Such large scale structures form via gravitational instability, and then
turbulent motions inside the structures as well as streaming motions
along the structures necessarily exist (\eg Kang \etal 1994).
Hence, magnetic fields in the early universe, if existed, should have
been modified and amplified by those flow motions (\eg Kulsrud \& Anderson
1992; Kulsrud \etal 1997).
As a result, we expect that {\it the ``geometry'' of cosmic magnetic
field, which means the spatial distribution of the field strength and its
orientation, should be correlated with the large-scale nonlinear
structures of the universe}.
In other words, the field strength increases with the matter density
and its orientation tends to align with the sheets and filaments, while
its random component is associated with the local turbulent motions.
The magnetic fields should be strong along the walls of cosmic bubbles,
even stronger along the filamentary superclusters, and the strongest
inside the clusters of the galaxies,
while they are very weak inside the voids.

Based on the proposition that the large scale magnetic field is
correlated with the large scale structure of the universe, we consider a
way to estimate the field strength which relies on the observational data.

The cosmic magnetic field together with free electrons in the
intergalactic medium (IGM) induces the Faraday rotation in polarized radio
waves from extra-galactic sources.
The observational RM data of quasars show a systematic growth of RM
with redshift, $z$, and limit RM to $\sim 5~{\rm rad~m^{-2}}$ or less at
$z=2.5$ (Kronberg \& Simard-Normandin 1976; Kronberg 1994 and
references therein).
Adopting a model for the distribution of the large scale magnetic field,
these data can be used to constrain the strength of the magnetic field.
For example, the upper limit estimated by Kronberg (1994) was based
on a model in which the field is uniform within the reversal scale,
the orientation is random, and there are $N=l_s/L_{rev}$ reversals along
the line of sight to a source $l_s$ Mpc away from us.

In the present paper, we re-derive the upper limit by taking a new model
in which the intergalactic magnetic field is mostly confined within
filaments and sheets but very weak inside voids, rather than the simple
model of uniform field.
Of course the field would be strongest inside clusters, but their
contribution is usually excluded in the observed RM.
If we take only the geometric consideration that the field distribution
along the lines of sight to radio sources has a small filling factor,
high only inside filaments and sheets but low inside voids,
then it is obvious that the expected magnetic field strength in
filaments and sheets should be higher than the value derived from the
uniform field model.
But this geometric model is too simple, since the electron distribution
as well as the field direction are uncertain.
Here, we take a more practical approach by adopting numerical data in a
simulated universe which includes a magnetic field, which is evolved during
the large scale structure formation. With the magnetic field distribution,
specific to this simulation model, we argue that the magnetic field strength in
filaments and sheets is limited to ${\bar B}_{fs} \la 1~\mu{\rm G}$.
We should emphasize that {\it this limit should be approximately valid
regardless of details of any model for the origin of the large scale magnetic
field, provided that it is correlated with the large scale structure of the
universe}.

In the next section, we describe the model simulation and the
resulting magnetic field geometry.
In \S3, we explain
the procedure to calculate the upper limit of the magnetic field
strength in filaments and sheets constrained by the observed RM.
In the final section, we discuss the implications of the possible
existence of $\sim 1~\mu{\rm G}$ or less magnetic field
in filaments and sheets.

\section{Model Simulation}

In this section we describe one particular simulation of the origin and growth
of magnetic fields, but will retain at the end only the geometry and the {\it
relative} strength of the magnetic fields across the cosmological structures. 
It is this structure for which we require a realistic simulation to probe the
universe for its Rotation Measure across a large number of simulated paths.

The generation of the large scale magnetic field by the Biermann battery
mechanism (L. Biermann 1950) and its subsequent evolution has been followed by
solving the equation
\begin{equation}
{\partial{\bf B}\over\partial t} = \nabla\times({\bf v}\times{\bf B})
+ {c\nabla p_e\times\nabla n_e\over n^2_e e}
\label{b_noncos}
\end{equation}
along with the cosmological hydrodynamic equations.
Here, $p_e$ is the electron pressure and $n_e$ is the electron number
density.
The simulation has been done with a version of the cosmological
hydrodynamic code which extends the one described in Ryu \etal (1993)
by incorporating the above equation ``passively''.
That is, the Lorenz force term in the momentum equation has been ignored.
Alternatively, the equation for the magnetic field could have been
incorporated by solving the full MHD equations.
Then the code would have been more diffusive.

A standard cold dark matter (CDM) model universe with total $\Omega=1$
has been simulated in a periodic box with $(32 h^{-1}{\rm Mpc})^3$
volume using $128^3$ cells and $64^3$ particles from $z_i=20$ to $z_f=0$.
The values of other parameters used are $\Omega_b=0.06$,
$h=1/2$, and $\sigma_8=1.05$.
A more detailed description of the simulation can be found in
in Kulsrud \etal (1997).

In the simulations which include the evolution of baryonic matter
as well as that of dark matter, accretion shocks form in the infalling
flows towards the growing nonlinear structures such as sheets,
filaments and clusters (\eg Kang \etal 1994; Ryu \& Kang 1997b).
We note that some evidence has been found in radio relic sources, that these
accretion shocks have actually been detected (En{\ss}lin \etal 1998).  The
properties of the shocks depend upon the power spectrum of the initial
perturbations on a given scale as well as the background expansion in a given
cosmological model. These shocks are the sites of the generation of weak seed
field by the battery mechanism represented by the last term in the right hand
side of Eq. (\ref{b_noncos}).
The evolution of the strength of the seed field generated was plotted in
Figure 1 of Kulsrud \etal (1997).
It grows monotonically to the mass averaged value of order
$ 10^{-21}{\rm G}$ by $z\sim2-3$, and then levels off without further
increase.
The saturation is believed to be due to the finite numerical
resistivity inherent in the numerical scheme used to solve Eq.
(\ref{b_noncos}).
At the saturation, the magnetic energy is much smaller (by $\sim10^{30}$)
than the kinetic energy.
So the dynamical influence of magnetic field into flow motion can be
ignored, and neglecting the Lorenz force in the momentum equation is
justified.

To illustrate the general characteristics of the field geometry in
relation to the matter distribution and flow velocity, a two-dimensional cut
of the simulated box is presented in Figure 1.
The plot shows a region of $32h^{-1}\times20h^{-1}({\rm Mpc})^2$ with
a thickness of $0.25 h^{-1}{\rm Mpc}$ at $z=0$.
The three panels display baryonic density contours, velocity vectors,
and magnetic field vectors.

%
\begin{figure*}
\vspace{-0.3truein}
\epsfysize=9in\epsfbox[-20 0 530 795]{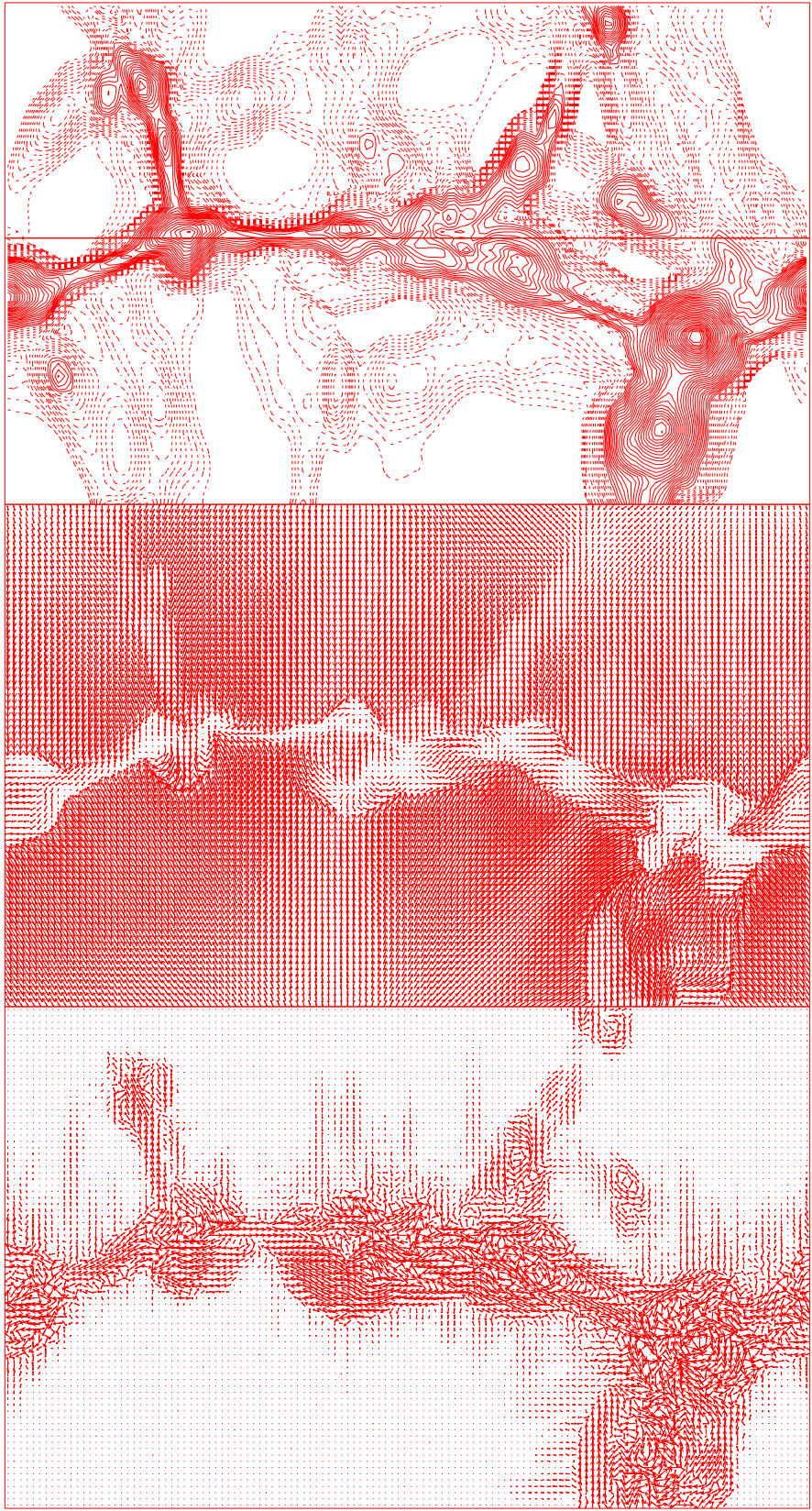}
\vspace{-0.3truein}
\caption[]{\label{fig_one}
Two-dimensional cut of the simulated universe at $z=0$.
The plot shows a region of $32h^{-1}\times20h^{-1}{\rm Mpc}^2$ with
a thickness of $0.25 h^{-1}{\rm Mpc}$, although the simulation was
done in a box of $(32 h^{-1}{\rm Mpc})^3$ volume.
The first panel shows baryonic density contours, the second panel
shows velocity vectors, and the third panel shows magnetic field vectors.
In the first panel, the solid lines contour the regions of
$\rho_b\geq{\bar\rho_b}$ and the dotted lines contour those of
$10^{-1.2}{\bar\rho_b}\leq\rho_b<{\bar\rho_b}$.
In the third panel, the vector length is proportional to the log of
magnetic field strength.
}
\end{figure*}

We can see accretion shocks around the high density structures of clusters,
filaments, and sheets in the density contours.
The locations of the shocks can be seen more clearly in the velocity
vector plot where accretion motions stop suddenly.
In the highest density regions of clusters, flow motion is turbulent.
Also we can see the streaming flow motion along the structures, that is,
the accretion motion towards clusters in filaments and the accretion
motion towards filaments in sheets.
On the top of the streaming motion, we expect some amount of turbulent
motion which is induced by the cooling, gravitational, and nonlinear thin
shell instabilities in filaments and sheets (Vishniac 1994;
Anninos, Norman \& Anninos 1995; Valinia \etal 1997).
So the magnetic field in filaments and sheets could have been stretched
by the streaming motion as well as amplified by the turbulent motion.
As a result, the amplified magnetic field in filaments and sheets is
expected to be aligned with the structures, as shown in Figure 1.

Here, we propose that {\it the geometry of the magnetic field in the
simulated universe, with an arbitrary normalization for the field
strength, represents the geometry of the real magnetic field correlated
with the large scale structure of the universe}.  Thus we take from the
simulation only the geometry of the field distribution and its relative
strength across the spatial distribution.

\section{Upper Limit on the Magnetic Field in Filaments and Sheets}

We have calculated the expected RMs induced by the magnetic field in
filaments and sheets in the simulated universe and compared them with
the observed upper limit of RM of distant quasars.

Due to the field reversal of randomly oriented magnetic field along lines
of sight, the RMs from sources at a given redshift have a Gaussian
distribution centered at zero and its standard deviation gives a statistical
measure for the field strength.
The standard deviation of the distribution of the RMs from quasars at
$z\sim 2.5$ is ${\rm RM} \la 5~{\rm rad}~{\rm m}^{-2}$ (Kronberg 1994).

The RM which measures the amount of the rotation is given by
\begin{equation}
{\rm RM} \equiv {\psi\over\lambda^2} =
{e^3\over2\pi m_e^2c^4}{\int_0^{l_s}n_eB_{\|}{\lambda(l)^2\over\lambda}dl}_,
\label{rm_noncos}
\end{equation}
where $\psi$ is the rotated angle, $\lambda(l)$ is the wavelength of
the polarized wave along the propagating path, $l$, $\lambda$ is the
observed wavelength, and $B_{\|}$ is the magnetic field component along
$l$ (\eg Shu 1991).
The RM for a source at a given redshift $z$ can be calculated numerically
by the following line integral
\begin{equation}
{\rm RM}\left({\rm rad}\over{\rm m}^2\right)
= 8.1\times10^5\int_0^{l_s}{\tilde n_e}B_{\|}dl
= 9.2\chi h^2{\int_0^{l_s}{\Omega_b}B_{\|}dl}_,
\label{rm_cosmo}
\end{equation}
where $l$ is the comoving length in the unit of Mpc, $B_{\|}$ is
the proper magnetic field in $\mu{\rm G}$, $\tilde n_e$ is the
comoving electron number density in ${\rm cm}^{-3}$,
$\Omega_b$ is the baryonic density as a fraction of the critical
density, and $\chi$ is the ionization fraction.
The integration length $l$ is related to the redshift by the following
equation
\begin{equation}
dl = -{c\over H_o}(1+z)^{-1}(1+\Omega_o z)^{-1/2}dz.
\label{dl_cosmo}
\end{equation}
In the calculation, we have assumed the gas inside filaments and sheets
is fully ionized (\ie $\chi=1$).
The simulated {\it distributions} of the gas and magnetic field at $z=0$,
0.5, 1, and 2 have been used, and they have been interpolated in between
to integrate the above equation up to $z=2.5$.

For the relative growth of the field strength for $2.5\ge z \ge 0$,
however, we need to adopt a model,
since the numerical simulation could not follow the growth
of magnetic field in detail due to numerical resistivity.
Again, the normalization of the field strength is arbitrary.
Considering that theoretical studies on dynamo processes on large
scales do not yet provide a unambiguous model, 
we assume for simplicity that the magnetic field has reached 
an ``equipartition'' with the energy in turbulence. 
Then its energy should be proportional to the kinetic energy which is,
in turn, proportional to the gravitational potential energy.
Then, the self-similar solution for an $\Omega=1$ universe gives
\begin{equation}
{\bar B} \propto (1+z)^{2(2n\epsilon-1)\over n\epsilon},
\label{bz}
\end{equation}
where $n=1$ is for sheets, 2 for filaments, and $3$ for clusters
(Fillmore \& Goldreich 1984).
Here, $\epsilon$ is the parameter which characterizes the initial
perturbation of the self-similar solution and has $0< \epsilon \leq 1$.
Hence, for filaments and sheets, we expect 
\begin{equation}
{\bar B}_{fs}\propto(1+z)^q
\end{equation}
with $0 \la q \la 1.5$.
However, we have also considered the model in which the magnetic field
grows as time passes (Welter, Perry, \& Kronberg 1984).
So, we have explored the cases of $-1.5\leq q \leq 1.5$.  This large 
range of the parameter $q$ explored here should encompass many models for the
origin of magnetic fields in the cosmos (See, \eg the discussions in L.
Biermann 1950; Parker 1958; Rees 1987; Beck \etal 1996; P. Biermann 1997;
Kronberg \& Lesch 1997).

With the above prescription for the evolution of magnetic field in
filaments and sheets, we have integrated Eq. (\ref{rm_cosmo})
up to $z=2.5$ along $3\times10^4$ randomly chosen paths.
Figure 2 shows the distribution of the expected RMs for the model with $q=0$.
The magnetic field is normalized so that the standard deviation of the
Gaussian fit (the solid line) to the expected RMs matches the observational
upper limit of ${\rm RM}=5~{\rm rad}~{\rm m}^{-2}$.
With this normalization for the field strength, the upper limits in the rms
of magnetic field in filaments and sheets can be calculated for several
models.
They are given in Table 1 for the several values of $q$, and they are
$\sim 1~\mu{\rm G}$ for $h=1/2$.

%
\begin{figure}
\vspace{-0.2truein}
\epsfysize=3.3in\epsfbox[105 210 480 580]{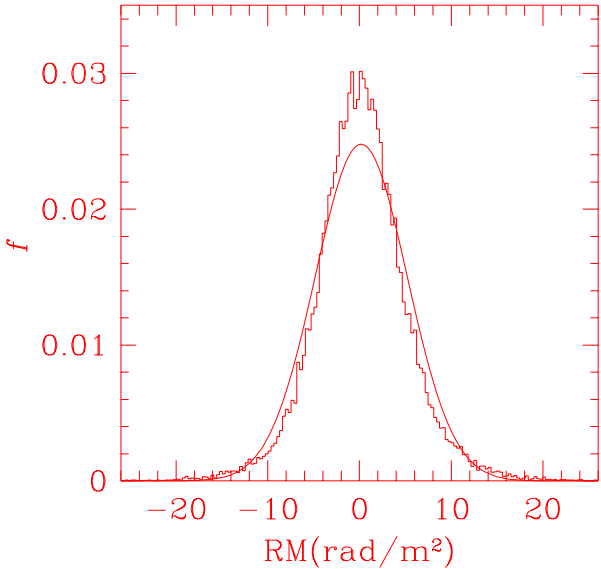}
\vspace{-0.1truein}
\caption[]{\label{fig_two}
Histogram of the fraction of the expected RM along
$3\times10^4$ randomly chosen paths up to $z=2.5$ in the simulated
universe.
The solid line represents the Gaussian fit.
The exponent in the temporal evolution of the rms of magnetic field in
filaments and sheets, $q$, is assumed to be zero.
The magnetic field strength is normalized so that the standard deviation
of the Gaussian fit matches the observational upper limit of
${\rm RM}=5~{\rm rad}~{\rm m}^{-2}$.
}
\end{figure}

\begin{table}
\caption{Upper Limits on the Magnetic Field Strength in Filaments
and Sheets}
\begin{tabular}{crc}
~ & $q$ $^{\rm a}$ & ${\bar B}_{z=0; fs}(\mu{\rm G})h_{0.5}^2$ $^{\rm b}$\\
~ & $1.5$  &  $0.27$ \\
~ & $1$    &  $0.42$ \\
~ & $0.5$  &  $0.63$ \\
~ & $0$    &  $0.89$ \\
~ & $-0.5$ &  $1.15$ \\
~ & $-1$   &  $1.39$ \\
~ & $-1.5$ &  $1.58$ \\
\end{tabular}

\medskip
$^{\rm a}$ The exponent in the temporal evolution of magnetic field
in filaments and sheets, ${\bar B}_{z=0; fs}\propto(1+z)^q$.

\smallskip
$^{\rm b}$ The rms of magnetic field in filaments and sheets at
the present epoch $(z=0)$ in units of $\mu{\rm G}$ for $h=1/2$.
\vspace{-0.05truein}
\end{table}

Figure 3 shows the distributions of gas density and magnetic field
strength along the line drawn in the first panel of Figure 1 for the
case with $q=0$.
We can see the field strength correlates well with the gas density and
ranges $0.2\la B\la 2 \mu{\rm G}h_{0.5}^{-2}$ in the regions of
filaments and sheets.

%
\begin{figure}
\epsfysize=3.2in\epsfbox[70 185 510 595]{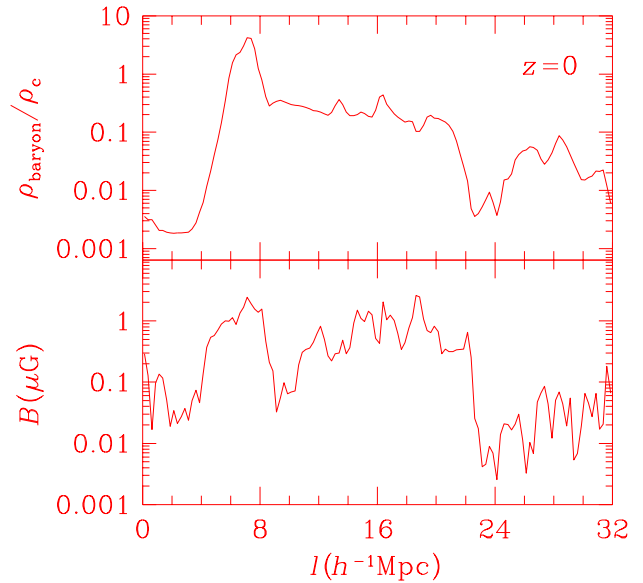}
\vspace{-0.1truein}
\caption[]{\label{fig_three}
Distributions of baryonic density and magnetic field strength at
$z=0$ along the line drawn in the first panel of Figure 1, when the
exponent in its temporal evolution is, $q$, zero.
The density is given in units of the critical density, while the
magnetic field is given in units of $\mu{\rm G}$ for $h=1/2$.
}
\end{figure}

These numbers clearly demonstrate that the essential result is nearly
independent of any of the details about the origin of the magnetic field, since
for the entire range of values tested in $q$ we obtain to within a factor of a
few the same result: The upper limit to the magnetic field in the sheet is
of order ${\bar B}_{fs} \la 1 \mu{\rm G}$.

\section{Conclusion and Discussion}

Our result indicates that, with the present value of the
observed limit in RM, the existence of magnetic field of up to
$\sim 1~\mu{\rm G}$ in filaments and sheets can not be ruled out,
if the cosmic magnetic field is preferentially distributed in these
nonlinear structures, rather than uniformly distributed in the
intergalactic space.  {\it We emphasize that this result is independent of the
details of how the magnetic field originated.}
However, it is not certain 
if the dynamo processes such as the one considered
in Kulsrud \etal (1997) can amplify a week seed field to
the above field strength.

Interestingly, there is  confirmation of such a magnetic field in one case.
Kim \etal (1989) reported the possible existence of a intercluster magnetic
field of $0.3-0.6~\mu{\rm G}$ in the plane of the Coma/Abell 1367 supercluster.
They reached the result from the observation using high dynamic range
synchrotron images at $327~{\rm MHz}$.
Clearly, this result should be confirmed in other regions of the sky with
the newly available possibilities of low radio frequency interferometry
such as with the Giant Meterwave Radio Telescope (GMRT) in India,
or the new receiver systems on the Very Large Array (VLA) in the USA.

We may compare the above upper limit with the strength of the magnetic
field whose energy is in equipartition with the thermal energy of the gas
in filaments and sheets.
The equipartition magnetic field strength can be written as
\begin{equation}
B = 0.33 h \sqrt{T \over 3 \times 10^6 {\rm
K}} \sqrt{\rho_b \over 0.3 \rho_c}~\mu{\rm G}. \label{b_equip}
\end{equation}
The fiducial values $T = 3 \times 10^6 {\rm K}$ and $\rho_b = 0.3 \rho_c$
may considered to be appropriate for filaments (see, Kang \etal 1994 and
Figure 3).
The values appropriate for sheets should be somewhat smaller.
So, our upper limit set by RM is close to, but several times larger than,
the equipartition magnetic field strength.

Filaments and sheets with the above fiducial temperature and gas density
can contribute radiation at X-ray wavelengths to the cosmological background.
Using the above values and a typical size of $10h^{-1}$Mpc, the luminosity
of filaments and sheets in the soft X-ray band ($0.5{-}2\,$keV) is
expected to be $\la 3\times 10^{41}{\rm erg~s^{-1}}$.
Actually, Soltan \etal (1996) and Miyaji \etal (1996) reported the
detection of extended X-ray emission from structures of a comparable size
and a luminosity of $\approx 2.5 \times 10^{43}{\rm erg~s^{-1}}$, which is
correlated to Abell clusters.
These observations suggest that the temperature and gas density
outside clusters could even be considerably larger than the fiducial
values assumed above.
In any case, if these observations can be further confirmed, and the
equipartition of magnetic field and gas energies is assumed, it would imply
the existence of a $\sim 1\mu{\rm G}$ magnetic field on large scales
outside galaxy clusters.

The possible existence of strong magnetic field of $1~\mu{\rm G}$ or less in
filaments and sheets has many astrophysical implications, some of which we
briefly outline in the following:

The large-scale accretion shocks where seed magnetic fields could be
generated are probably the biggest shocks in the universe with a typical
size $\ga$ a few (1-10) Mpc and very strong with a typical
accretion velocity $\ga$ a few $1000~{\rm km}~{\rm s}^{-1}$.
The accretion velocity onto the shocks around clusters of a given
temperature, or a give mass to radius ratio $M_{cl}/R_{cl}$, is smaller
in a universe with smaller $\Omega_o$, and is given as $v_{acc}
\approx 0.9-1.1 \times 10^3~{\rm km~s^{-1}}
[(M_{cl}/R_{cl})/(4 \times 10^{14}{\rm M}_{\odot}/{\rm Mpc})]^{1/2}$
in model universes with $0.1 \le \Omega_o \le 1$ (Ryu \& Kang 1997a).
With up to $\sim 1~\mu{\rm G}$ or less magnetic field around them,
the large-scale accretion shocks could serve
as possible sites for the acceleration of high energy cosmic rays by
the first-order Fermi process (Kang, Ryu \& Jones 1996;
Kang, Rachen \& Biermann 1997).
Although the shocks around clusters would be the most efficient sites for
acceleration, those around filaments and sheets could make a significant
contribution as well (Norman, Melrose \& Achterberg 1995).
With the particle diffusion model in quasi-perpendicular shocks
(Jokipii 1987), the observed cosmic ray spectrum near
$10^{19}{\rm eV}$ could be explained with reasonable parameters if about
$10^{-4}$ of the infalling kinetic energy can be injected into the
intergalactic space as the high energy particles (if an $E^{-2}$ spectrum of
cosmic rays is assumed; for a slightly steeper spectrum as suggested by radio
relic sources, the efficiency is closer to 0.1, En{\ss}lin \etal 1998).

The discoveries of several reliable events of high energy cosmic rays at
an energy above $10^{20} {\rm eV}$ raise questions about their origin and
path in the universe (a recent review is P. Biermann 1997), since their
interaction with the cosmic microwave background radiation limits the
distances to their sources to less than 100 Mpc, perhaps within our Local
Supercluster. The Haverah Park and Akeno data indicate that their arrival
directions are in some degree correlated with the direction of the
Supergalactic plane (Stanev \etal 1995; Hayashida \etal 1996; Uchihori \etal
1996). In Biermann, Kang \& Ryu (1996), we noted that
if the magnetic field of $\sim 1\mu{\rm G}$ or less exists inside
our Local Supercluster and there exist accretion flows infalling toward
the supergalactic
plane, it is possible that the high energy cosmic rays above the so-called
GZK cutoff ($E> 5\times 10^{19}$ eV) can be confined
to the supergalactic plane sheet, an effect
analogous to solar wind modulation.  In each case a shock front pushes
energetic particles upstream as seen from its flow. Obviously, this effect
would occur only for a small part of phase space, namely those particles with
a sufficiently small initial momentum transverse to the sheet.
This would explain naturally the correlation between the arrival direction
of the high energy cosmic rays and the supergalactic plane. Also, confinement
means that for all the particles captured into the sheets, the dilution with
distance $d$ is $1/d$ instead of $1/d^2$, increasing the cosmic ray flux
from any source appreciably with respect to the three-dimensional dilution.
So we may see sources to much larger distances than expected so far.  On the
other hand, particles with a larger initial momentum transverse to the sheet
would be strongly scattered, obliterating all source information from their
arrival direction at Earth.

With the magnetic field in the intergalactic medium, charged particles
passing it would not only experience deflection.
It would also smear out their arrival direction as well as delay
their arrival time.
Plaga (1995) suggested that an exhibition of this latter effect would be the
delay of the arrival times of $\gamma$-rays from a cascade caused by
photons from highly time-variable extragalactic sources.
However, with strong fields of $\sim 1~\mu{\rm G}$ or less in filaments
and sheets intervening between the sources and us, the expected consequence
would be a strong smearing of the sources rather than the delay of arrival
times (Kronberg 1995).
But, for details, calculations following the propagation of photons and
charged particles should be done.

Another interesting exploration is to study the radiation emission arising
from filaments and sheets strong magnetic field of $1~\mu{\rm G}$ or less.
We noted that the original estimate of the magnetic field in the plane of
the Coma/Abell 1367 supercluster was based on a synchrotron radio continuum
measurement (Kim \etal 1989).
We also noted that the thermal Bremsstrahlung emission in the soft X-ray
band  may provide an interesting check on the work presented here.
Whether these structures can contribute at other wavelengths, such as
$\gamma$-ray energies, to the cosmological background remains an an
unanswered but challenging question at this time.

These issues will be discussed elsewhere.

\section{acknowledgments}

We are grateful to Dr. P.P. Kronberg for extensive comments on the manuscript.
We also thank Drs. R. Beck and J. Rachen for comments. PLB wishes to thank Drs.
T. En{\ss}lin, R. Protheroe, G. Sigl, and T. Stanev for extensive discussions
of cosmological magnetic fields. The work by DR and HK was supported in part by
NASA HPCC/ESS at the University of Washington. The work by DR was supported
in part by Seoam Scholarship Foundation and by KOSEF through the 1997
Korea-US Cooperative Science Program 975-0200-006-2 at Chungnam National
University.

\end{document}